# Kullback–Leibler Divergence as a Measure of Irreversible Information Loss Near Black Hole Horizons

Tatsuaki Tsuruyama

Tatsuaki Tsuruyama[1,2,3]
[1]Department of Physics, Tohoku University, Sendai 980-8578, Japan
[2]Department of Drug Discovery Medicine, Kyoto University, Kyoto 606-8501, Japan
[3]Department of Clinical Laboratory, Kyoto Tachibana University, Kyoto 607-8175, Japan
e-mails: tsuruyam@ddm.med.kyoto-u.ac.jp

**Abstract**
We present a unified theoretical framework that integrates information theory, thermodynamics, and general relativity to analyze the fundamental limit of decoding time-encoded signals in curved spacetime. In particular, we introduce the Kullback–Leiblerdivergence (KLD) as a quantitative measure of the mismatch between the transmitted and received symbol distributions inducedby gravitational time dilation. Using a minimal communication model, we derive the critical radius at which information decodingbecomes thermodynamically impossible due to the divergence of the KLD. We show that this radius approaches the Schwarzschildhorizon in the limit where the information entropy cost becomes negligible relative to the transmission energy. This result provides anovel information-theoretic interpretation of the event horizon as a boundary of irreversible information loss governed by universalthermodynamic principles. Our framework offers new insights into the entropic and energetic constraints on communication instrong gravitational fields and may extend to general relativistic and quantum information settings.

**1 Introduction**

Understanding the limits and efficiency of information transmission is a foundational issue in information theory and thermodynamics [1–6]. Recent developments in information thermodynamics have revealed deep connections between entropy production, energy dissipation, and statistical fluctuations in non-equilibrium systems, formalized in frameworks such as the thermodynamic uncertainty relation [7–10].

In this study, we focus on a specific but universal measure of information distortion, the Kullback–Leibler divergence (KLD), as a diagnostic tool for assessing the stability of information transmission under relativistic effects. When signals are emitted from stronggravitational environments such as black hole neighborhoods, gravitational time dilation introduces temporal distortions betweenthe transmitted and received symbol durations, resulting in mismatches between the respective probability distributions of codelengths [11,12]. To investigate the implications of these distortions, we construct a unified theoretical framework that integrates information theory, thermodynamics, and general relativity within an information geometric setting. Central to this analysis is the KLD, whichquantifies the distinguishability between the sender's original distribution and the receiver's distorted interpretation [20]. Importantly, the KLD also plays a fundamental role in the information geometry of statistical manifolds [13,14]. In this geometric formulation, probability distributions are regarded as points on a differential manifold, and KLD serves as the generating functionfor the Fisher information metric and dual affine connections that define its geometry. The divergence structure naturally endowsthe space of code distributions with curvature, and this curvature becomes pronounced in relativistic contexts where time distortionacts as a geometric perturbation on the manifold of possible symbol structures.

We show that the KLD diverges as the transmission of the code sequence approaches the Schwarzschild radius, reflecting a fundamental instability in the decoding process. This approach yields a new interpretation of black hole entropy, not as a statisticalproperty of microstates, but as the maximal rate of recoverable information constrained by geometric and thermodynamic principles [15,16]. Through this KLD-centered approach based upon information geometry, we offer a novel and physically perspective on how spacetime curvature governs the fidelity of information and attempt a more unified understanding of entropy, irreversibility, and thefundamental limits on signal decodability in gravitational fields.

**2.1 Model setup**

We consider a communication scenario in which a sender encodes a message into a sequence of optical pulses, each representing a symbol $x_j$. Each symbol is characterized by a code length $\tau_j^{(a)}$, which determines the duration of the optical pulse, and an associated probability $p_j$ representing the statistical structure of the message. The sequence is transmitted through a curved spacetime region, such as near a black hole. A receiver located at a distant position observes the signal and attempts to decode the original symbol sequence. However, due to gravitational time dilation, the receiver perceives different time durations $\tau_j^{(b)}$ for each symbol, resulting in a mismatch between the transmitted and received code structures. Let the total number of transmitted symbols be denoted by $X$. The average code length at the sender's side is defined as:

$$\langle\tau\rangle = \sum_{j=1}^{n} p_j \tau_j \quad (1)$$

where $n$ is the number of distinct symbol types. The total transmission time from the sender's perspective is then:

$$T = X \langle\tau\rangle \quad (2)$$

Assuming each symbol is realized by an optical pulse emitted with constant power $P$, the energy cost to transmit symbol $x_j$ is:

$$\varepsilon_j = P \cdot \tau_j \quad (3)$$

yielding a total transmission energy:

$$E = X \sum_{j=1}^{n} p_j = PT \quad (4)$$

To maximize the Shannon entropy of the message distribution under a fixed transmission time constraint, we impose:

$$\sum_{j=1}^{n} p_j \tau_j = \langle \tau \rangle \tag{5}$$

$$\sum_{j=1}^{n} p_j = 1 \tag{6}$$

Solving the variational problem yields an exponential distribution (Appendix A):

$$p_j = \frac{e^{-\beta \tau_j}}{Z}, \quad Z = \sum_{j=1}^{n} e^{-\beta \tau_j} \tag{7}$$

with β controlling the sharpness of the distribution. Based on this, the maximal entropy becomes:

$$S_{\max} = \beta \langle \tau \rangle \tag{8}$$

And we define the entropy rate as:

$$\sigma = \frac{S_{\max}}{\langle \tau \rangle} = \beta \tag{9}$$

and the energy cost can then be expressed as:

$$E = P \langle \tau \rangle \tag{10}$$

and the energy cost per unit entropy as:

$$\frac{dE}{dS_{\max}} = \frac{P}{\sigma} = T_{\text{info}}, \tag{11}$$

where $T_{\text{info}}$ is the information *thermodynamic temperature*.

## 2.2 Sender and receiver 4-velocities in stationary spacetimes

We consider a static, spherically symmetric spacetime such as Schwarzschild geometry, described by the line element:

$$ds^2 = -f(r)c^2 dt^2 + \frac{1}{f(r)} dr^2 + r^2 d\Omega^2 \tag{12}$$

where

$$f(r) = 1 - \frac{2GM}{c^2 r} \tag{13}$$

This geometry admits a global time-like Killing vector field:

$$\xi^\mu = (\partial_t)^\mu \tag{14}$$

reflecting time-translation symmetry. Observers whose 4-velocity is aligned with ξμ are said to be stationary. Their normalized
4-velocity is given by:

$$u^\mu = \frac{\xi^\mu}{\sqrt{-\xi^\nu \xi_\nu}} = \left(\frac{1}{\sqrt{f(r)}}, 0, 0, 0\right) \quad (15)$$

using $g_{tt} = -f(r)$
.

For a sender at radius $r_a$ and a receiver at $r_b$, their 4-velocities are:

$$u_a^\mu = \left(\frac{1}{\sqrt{f(r_a)}}, 0, 0, 0\right) \quad (16)$$

$$u_b^\mu = \left(\frac{1}{\sqrt{f(r_b)}}, 0, 0, 0\right) \quad (17)$$

We now consider an optical pulse emitted from the sender and received by the receiver, traveling along a radial null geodesic γab.
The null wavevector $k_\mu$ satisfies:

$$g_{\mu\nu} k^\mu k^\nu = 0, \quad k^\mu \nabla_\mu k^\nu = 0 \quad (18)$$

Assuming radial propagation, the wavevector can be written as:
$$k^\mu = \omega(1, f(r), 0, 0) \quad (19)$$

where $\omega$ is the conserved frequency as measured by an observer at infinity. The proper time duration τ of a pulse measured by a static observer is proportional to the projection:
$$\tau \propto -u^\mu k_\mu \quad (20)$$

Therefore, the durations at the sender and receiver locations are:
$$\tau_j^{(a)} \propto -u_a^\mu k_\mu \quad (21)$$

$$\tau_j^{(b)} \propto -u_b^\mu k_\mu \quad (22)$$

Taking the ratio, we obtain the gravitational time dilation factor:
$$\frac{\tau_j^{(b)}}{\tau_j^{(a)}} = \frac{u_b^\mu k_\mu}{u_a^\mu k_\mu} = \sqrt{\frac{f(r_a)}{f(r_b)}} \quad (23)$$

In the limit $r_b \to \infty$, w e h a v e f(rb)→1, and thus:

$$\tau_j^{(b)} = \tau_j^{(a)} \sqrt{1 - \frac{2GM}{c^2 r}} \quad (24)$$

which is the standard gravitational redshift relation.

## 2.3 Entropy and divergence under time dilation

Having established how code durations transform between sender and receiver frames via geometric projections, we now analyze the impact of this transformation on the statistical structure of the encoded message. The sender constructs a maximum entropy distribution based on the local durations $\tau_j^{(a)}$, with the form:

$$p_j^{(a)} = \frac{e^{-\beta \tau_j^{(a)}}}{Z^{(a)}}, \quad S^{(a)} = -\sum_j p_j^{(a)} \log p_j^{(a)} \tag{25}$$

However, due to gravitational time dilation, the receiver observes a distorted sequence of durations $\tau_j^{(b)}$, leading to an effectively different probability distribution:

$$p_j^{(b)} = \frac{e^{-\beta \tau_j^{(b)}}}{Z^{(b)}} \tag{26}$$

Although the KLD is formally coordinate invariant, its operational meaning depends on the information structure available to each observer. Here, we interpret:

$$D_{KL}(p^{(b)} \| p^{(a)}) = \sum_j p_j^{(b)} \log\left(\frac{p_j^{(b)}}{p_j^{(a)}}\right) \tag{27}$$

as the thermodynamic cost of restoring the distorted distribution $p_j^{(b)}$ to the original $p_j^{(a)}$, i.e., the irreversibility induced by spacetime curvature on the decoding process. Substituting the exponential forms of $p_j^{(a)}$ and $p_j^{(b)}$.

$$D_{KL}(r) = \beta \sum_j p_j^{(b)} \left(\tau_j^{(a)} - \tau_j^{(b)}\right) + \log\left(\frac{Z^{(a)}}{Z^{(b)}}\right) \tag{28}$$

From Eq. (22), we have:

$$\tau_j^{(a)} - \tau_j^{(b)} = \tau_j^{(a)} \left(1 - \frac{1}{\sqrt{1 - \frac{2GM}{c^2 r}}}\right) \tag{29}$$

Although Eq. (28) includes the partition function ratio $\log(Z^{(a)}/Z^{(b)})$, we neglect this term in the following approximation for the following reasons. First, under the assumption that the distribution $p_j^{(b)}$ is sharply peaked, the dominant contribution to the divergence arises from the mean shift in durations due to gravitational time dilation:

$$\sum_j p_j^{(b)} \left(\tau_j^{(a)} - \tau_j^{(b)}\right) \approx \beta \left(\frac{1}{\sqrt{1 - \frac{2GM}{c^2 r}}} - 1\right) \langle \tau^{(a)} \rangle \tag{30}$$

Second, although the partition functions Z(a) and Z(b) differ due to the rescaling of durations under time dilation, their ratio changes subdominantly compared to the leading divergence. Specifically, we have:

$$Z^{(a)} = \sum_j e^{-\beta \tau_j^{(a)}} \tag{31}$$

$$Z^{(b)} = \sum_j e^{-\beta \tau_j^{(a)}/\sqrt{1-\frac{2GM}{c^2 r}}} \tag{32}$$

The difference in their logarithms remains finite even as $r \to r_s$, while the duration-based term diverges. Therefore, the logarithmic term $\log(Z^{(a)}/Z^{(b)})$ can be safely neglected when analyzing the asymptotic behavior of $D_{KL}(r)$ near the Schwarzschild radius. This approximation simplifies the expression without altering its leading-order divergence structure. In this case, neglecting the subdominant term $\log(Z^{(a)}/Z^{(b)})$, the divergence simplifies to:

$$D_{KL}(r) \approx \beta \left( \frac{1}{\sqrt{1-\frac{2GM}{c^2 r}}} - 1 \right) \tau^{(a)} \tag{33}$$

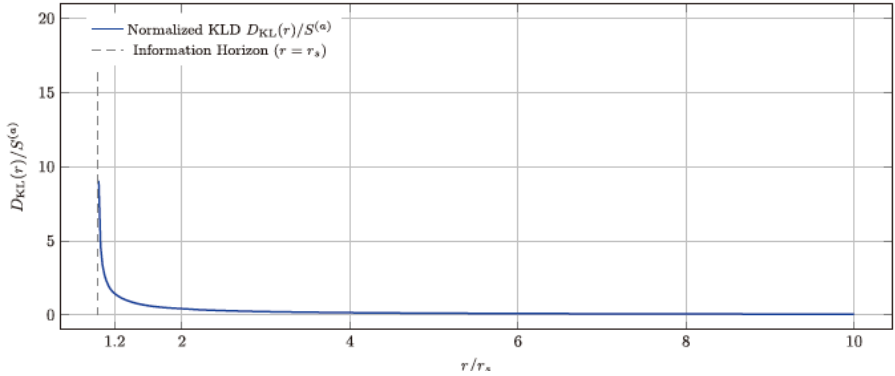

**Fig. 1** The normalized KLD $D_{KL}(r)/S^{(a)}$ diverges as the emitter approaches the Schwarzschild radius $r_s$. This divergence defines the decoding boundary, where gravitational time dilation renders the received distribution irrecoverable. The plot shows the function $(1/(1-x)^{1/2}-1)$ over the range $x=r/r_s \in [1.01, 10]$ using $S^{(a)}=1$.

Recalling the entropy expression $S(a) = \beta \tau^{(a)}$, we write:

$$D_{KL}(r) \approx \left( \frac{1}{\sqrt{1-\frac{2GM}{c^2 r}}} - 1 \right) S^{(a)} \tag{34}$$

To extract the asymptotic behavior as $r - r_s = 2GM/c^2$, define, $\varepsilon := r - r_s \ll r_s$, and expand:

$$1 - \frac{2GM}{c^2 r} \approx \frac{\varepsilon}{r_s} \quad \Rightarrow \quad \frac{1}{\sqrt{1 - \frac{2GM}{c^2 r}}} \sim \left(\frac{r_s}{\varepsilon}\right)^{1/2}$$

$$D_{KL}(r) \sim \left(\frac{r_s}{r - r_s}\right)^{1/2} S^{(a)}$$

(35)

This divergence reflects a fundamental breakdown of decodability near the Schwarzschild horizon: The receiver's effective symbol distribution becomes so distorted that restoring the original message requires infinite information-theoretic cost. In this limit, the curvature of spacetime enforces an irreversible deformation of the message ensemble, marking a boundary beyond which communication loses thermodynamic feasibility (Fig. 1).

**2.4 Free energy and the critical radius for decodability**

Having established the form of the KLD $D_{KL}(r)$ induced by gravitational time dilation, we now turn to its thermodynamic implications within the framework of nonequilibrium information thermodynamics. In this context, the generalized free energy quantifies the maximum amount of usable information (or extractable work) that can be recovered from an encoded signal, once entropic and distortion-induced losses are accounted for. Specifically, under gravitationalredshift, we interpret the KLD as a measure of irreversible entropy production that degrades decodability. We define an effective free energy available to the receiver as:

$$F^{(b)}(r) = \langle \varepsilon \rangle - T_{\text{info}}\left(S^{(a)} + D_{KL}(r)\right)$$

(36)

where
- $\langle \varepsilon \rangle$: average energy per symbol delivered to the receiver,
- $S^{(a)}$: Shannon entropy of the transmitted symbol distribution,
- $D_{KL}(r)$: additional entropy cost from time dilation,
- $T_{\text{info}}$: effective thermodynamic temperature of information encoding.

This free energy quantifies the receiver's capacity to decode the message: Decoding is thermodynamically feasible only when $F^{(b)}(r_{\text{crit}}) \geq 0$. The critical decoding radius is therefore defined by the conclusion:

$$F_{\text{crit}}^{(b)}(r) = 0 \qquad (37)$$

i.e., when the entire energetic resource is just sufficient to support both intrinsic and distortion-induced entropy. Substituting Eq. (36) into Eq. (37), we obtain the critical condition:

$$D_{KL}(r_{\text{crit}}) = \frac{\langle \varepsilon \rangle}{T_{\text{info}}} - S^{(a)} \qquad (38)$$

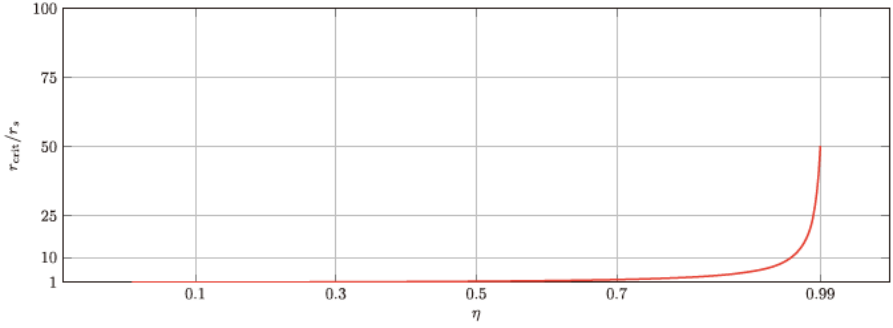

**Fig. 2** The critical decoding radius $r_{crit}$ diverges as the efficiency $\eta$ approaches unity. This identifies the Schwarzschild radius as the thermodynamic decoding limit in the low-entropy regime. The plot shows the function $r_{crit}/r_s = 1/(1-\eta^2)$ over the range $\eta \in [0.01, 0.99]$

Using the approximate form of DKL derived earlier [Eq. (34)], we write:

$$\left(\frac{1}{\sqrt{1-\frac{2GM}{c^2 r_{\text{crit}}}}} - 1\right) S^{(a)} = \frac{\langle \varepsilon \rangle}{T_{\text{info}}} - S^{(a)} \qquad (39)$$

Solving for the redshift factor, we rearrange:

$$\frac{1}{\sqrt{1-\frac{2GM}{c^2 r_{\text{crit}}}}} = \frac{\langle \varepsilon \rangle}{T_{\text{info}} S^{(a)}} \qquad (40)$$

Squaring both sides yields:

$$1 - \frac{2GM}{c^2 r_{\text{crit}}} = \left(\frac{T_{\text{info}} S^{(a)}}{\langle \varepsilon \rangle}\right)^2 \qquad (41)$$

Finally, solving for rcrit, we obtain:

$$r_{\text{crit}} = \frac{2GM}{c^2}\left(1 - \left(\frac{T_{\text{info}} S^{(a)}}{\langle \varepsilon \rangle}\right)^2\right)^{-1} \qquad (42)$$

To interpret this result, we introduce the decoding efficiency parameter $\eta \in (0, 1]$, defined by:

$$\langle \varepsilon \rangle = \eta P \langle \tau^{(a)} \rangle, \quad T_{\text{info}} S^{(a)} = P \langle \tau^{(a)} \rangle \qquad (43)$$

which implies:

$$\frac{T_{\text{info}} S^{(a)}}{\langle \varepsilon \rangle} = \frac{1}{\eta} \qquad (44)$$

Substituting into Eq. (42), we find the closed-form expression for the critical decoding radius:

$$r_{\text{crit}} = \frac{2GM}{c^2(1-\eta^2)} \tag{45}$$

This result admits two important physical limits:
• As decoding efficiency improves ($\eta \to 1$), the critical radius diverges: $\eta \to 1 \Rightarrow r_{crit} \to \infty$.
In this regime, even small gravitational redshifts induce thermodynamic instability in decoding.
• As decoding efficiency becomes poor ($\eta \ll 1$), we recover:
$$\eta \to 0 \Rightarrow r_{\text{crit}} \to \frac{2GM}{c^2} = r_{\text{s}},$$
recovering the classical notion of the Schwarzschild radius as the ultimate boundary of information recovery.

This formulation shows that decoding failure near a black hole is not merely geometric, but results from a quantifiable trade-off between signal energy and entropy distortion. The Schwarzschild horizon emerges as a limiting case of vanishing free energy for decoding, governed by the information-to-energy efficiency of the communication process.

**3 Discussion**

In this study, we constructed a theoretical framework for analyzing the stability of time-coded information transmission in curved spacetime from the perspective of information geometry. In particular, by analyzing the transmission of time-encoded information in curved spacetime, we showed that gravitational time dilation induces a systematic mismatch between the transmitted and received code structures. Herein, we developed a geometric formalism, in which the code durations are determined by the contraction of null wavevectors with observer 4-velocities aligned with Killing fields. This allows an observer-independent definition of gravitational time distortion and leads to a rigorous interpretation of signal timing. From a thermodynamic standpoint, the instability is characterized by the vanishing of the effective decoding free energy, which we defined as the net energy available after accounting for the entropic cost of distortion. This construction is consistent with the framework in information thermodynamics [6,7], where the KLD governs irreversible entropy production and limits extractable work.

Our result identifies a critical decoding radius $r_{crit}$, which exceeds the Schwarzschild radius in general and converges to it in the limit where the transmitted entropy cost becomes negligible relative to the energy budget. This result was obtained analytically in Eq. (45) and visualized in Figure 2, showing its divergence as the entropy-to-energy ratio approaches unity. It has been not noted that the KLD loss should be adopted as a fundamental measure of itself. However, by deriving the KLD near the Schwarzschild radius within a free energy framework, we found that it served as a natural and analytically tractable indicator in our geometric and thermodynamic formulation. This approach provided a promising means to elucidate the significance of information thermodynamics.

Thus, the event horizon can be reinterpreted as the boundary beyond which time-encoded information becomes fundamentally undecodable due to gravitational effects.

Unlike traditional formulations based on quantum field theory in curved spacetime [15,16], which interpret black hole entropy as a statistical property of inaccessible degrees of freedom, our approach treats entropy as a communication-theoretic quantityconstrained by energy and curvature. Figure 1 shows that $D_{KL}(r)$ diverges as $r \to r_s$, while Figure 2 quantifies how the critical decoding radius grows rapidly as the effective entropy-to-energy ratio $\alpha = T_{info} S^{(a)} / \langle \varepsilon \rangle$ approaches unity. These plots confirm the physical interpretation of the event horizon as an information-theoretic singularity. Furthermore, we introduced a unified framework by the identification of the critical decoding condition based on information free energy, in which

•energy conservation (via the transmission cost $\langle \varepsilon \rangle$),

•entropy generation (via the KLD),

•and causal structure (via the gravitational geometry)

are coherently integrated.

Our presented framework offers a general recipe for deriving information-theoretic limits in gravitational environments. For instance, in rotating spacetimes (e.g., Kerr geometry) [21], the frame dragging effect could couple symbol encoding rates with angular momentum, modifying the decoding condition. Moreover, in quantum information contexts, one could replace classicalKLD with quantum relative entropy or Holevo information, potentially defining a quantum decoding horizon. Future directions include theoretically extending the analysis to rotating and dynamical spacetimes and incorporating quantum information structures to further explore fundamental limits to information transmission in more general spacetime geometries [17–19].

## 4.Conclusion
In this work, we established a unified framework that integrates information theory, thermodynamics, and general relativity to analyze the fundamental limits of information decoding in curved spacetime. We introduced a minimal communication model in which messages are encoded into time-varying optical pulses and transmitted through a gravitational field. Gravitational time dilation causes distortions in the code durations, which manifest as statisticalmismatches in the received symbol distribution. By quantifying this distortion using the KLD, and by formulating the decoding condition in terms of an information-theoretic free energy, we derived an explicit expression for the critical decoding radius:

$$S = -\sum_j p_j \log p_j \tag{46}$$

This equation describes the boundary beyond which information becomes irreversibly undecodable due to curvature-induced entropy production. It provides a novel interpretation of the event horizon—not as a statistical artifact of quantum microstates, but as aphysically defined limit of information recoverability.

To achieve this, we developed a geometric formulation based on Killing vector-aligned 4-velocities and null geodesic transport, which allowed the durations and distributions to be defined in a coordinate-invariant and physically meaningful way. Our results demonstrate that gravitationally induced information distortion has a thermodynamic cost, and that decoding becomes energetically impossible once this cost exceeds the available transmission energy. This approach offers a promising direction for addressing long-standing questions about information loss and entropy generation near black holes, independently of the quantum field-theoretic microstate paradigm. It also opens the possibility of generalizing the concept of decoding horizons to rotating (Kerr) or dynamic spacetimes and to quantum information frameworks based on relative entropy or Holevo information. More broadly, we suggest that geometric limits on information recovery expressed via curvature-induced divergence in statistical structure may represent a general organizing principle at the intersection of spacetime geometry and information theory.

**Appendix**

Appendix A: Derivation of the exponential distribution To derive the maximum entropy distribution under the constraint of fixed average duration $\langle\tau\rangle$, we maximize the Shannon entropy:

$$\sum_j p_j \tau_j = \langle\tau\rangle, \quad \sum_j p_j = 1 \tag{47}$$

subject to

$$\sum_j p_j \tau_j = \langle\tau\rangle \tag{48}$$

$$\sum_j p_j = 1 \tag{49}$$

Introducing Lagrange multipliers α and β, we define the Lagrangian:

$$\mathcal{L} = -\sum_j p_j \log p_j - \alpha\left(\sum_j p_j - 1\right) - \beta\left(\sum_j p_j \tau_j - \langle\tau\rangle\right) \tag{50}$$

Taking the derivative with respect to $p_j$ and setting to zero gives:

$$\frac{\partial \mathcal{L}}{\partial p_j} = -(1 + \log p_j) - \alpha - \beta\tau_j = 0 \tag{51}$$

$$\log p_j = -1 - \alpha - \beta\tau_j \Rightarrow p_j \propto e^{-\beta\tau_j} \tag{52}$$

which yields:

$$\frac{\partial \mathcal{L}}{\partial p_j} = -(1 + \log p_j) - \alpha - \beta\tau_j = 0 \tag{53}$$

$$\log p_j = -1 - \alpha - \beta \tau_j \Rightarrow p_j \propto e^{-\beta \tau_j} \tag{54}$$

Normalizing this distribution gives:

$$p_j = \frac{e^{-\beta \tau_j}}{Z}, \quad Z = \sum_j e^{-\beta \tau_j} \tag{55}$$

**References**
1. C.E. Shannon, Bell Syst. Tech. J. 27, 379 (1948)
2. T.M. Cover, J.A. Thomas, Elements of Information Theory, 2nd edn. (Wiley, London, 2006)
3. R. Landauer, IBM J. Res. Dev. 5, 183 (1961)
4. E.T. Jaynes, Phys. Rev. 106, 620 (1957)
5. C.H. Bennett, Int. J. Theor. Phys. 21, 905 (1982)
6. J.M. Horowitz, M. Esposito, Phys. Rev. X 4, 031015 (2014)
7. J. Parrondo, J.M. Horowitz, T. Sagawa, Nat. Phys. 11, 131 (2015)
8. T. Sagawa, M. Ueda, Phys. Rev. E 85, 021104 (2012)
9. A.C. Barato, U. Seifert, Phys. Rev. Lett. 114, 158101 (2015)
10. A. Dechant, J. Phys. A Math. Theor. 52, 035001 (2018)
11. W.G. Unruh, Phys. Rev. D 14, 870 (1976)
12. R.M. Wald, Quantum Field Theory in Curved Spacetime and Black Hole Thermodynamics (University of Chicago Press, Chicago, 1994)

13. S. Amari, H. Nagaoka, Methods of information geometry. Translations of Mathematical Monographs, V ol. 191 (American Mathematical Society and Oxford University Press, Providence, RI, 2000) originally published in Japanese (1993)
14. N. Ay, J. Jost, H. Minh, Le, L. Schwachhöfer, Information Geometry, Ergebnisse der Mathematik und ihrer Grenzgebiete , V ol. 64 (Springer, 2017).
15. J.D. Bekenstein, Phys. Rev. D 7, 2333 (1973)
16. S.W. Hawking, Commun. Math. Phys. 43, 199 (1975)
17. J.M. Bardeen, W.H. Press, S.A. Teukolsky, Astrophys. J. 178, 347 (1972)
18. V.P. Frolov, I.D. Novikov, Black Hole Physics: Basic Concepts and New Developments (Springer, Berlin, 1998)
19. I. Fuentes-Schuller, R.B. Mann, Phys. Rev. Lett. 95, 120404 (2005)
20. T. Tsuruyama, Entropy 20, 438 (2018)
21. R.P. Kerr, Phys. Rev. Lett. 11, 237 (1963)
22. H.S. Alexander, Probl. Info. Transm. 9, 177 (1973)